\documentstyle[12pt,axodraw,epsfig]{article}

\renewcommand{\title}{        New Insight in Dynamical Symmetry Breaking
\\                                   via Imaginary Part Analysis
}
\newcommand{\autha}{                     Andreas Blumhofer
}
\newcommand{\authb}{                      Johannes Manus
}

\advance\textheight by \topskip
\renewcommand{\baselinestretch}{1.2}
\renewcommand{\thefootnote}{\fnsymbol{footnote}}
\renewcommand{\*}[1]{\mbox{$ #1 $}}
\newcommand{\gapprox}{\begin{array}{c} > \\[-0.43cm] \sim \end{array}}
\newcommand{\beq}{\begin{equation}}
\newcommand{\eeq}{\end{equation}}
\newcommand{\bea}{\begin{eqnarray}}
\newcommand{\eea}{\end{eqnarray}}
\def\slash#1{#1\!\!\!/\!\,\,}

\newcommand{\I}{\Im m}
\newcommand{\R}{\Re e}
\renewcommand{\l}{\ell}

\newcommand{\bild}[3]{
  \begin{minipage}[t]{7.5cm}
  \begin{center}\begin{picture}(200,200)
  \epsfig{file=#1.ps,height=4cm,width=4cm,
          bbllx=3.2cm ,bblly=3.2cm ,bburx=11.5cm ,bbury=11.5cm ,angle=0}
  \end{picture}\end{center}
  \vspace*{-8cm}
  \begin{center}\begin{picture}(200,200)
    #2
  \end{picture}\end{center}
  \begin{fig} #3 \label{#1} \end{fig}
  \end{minipage}}

\begin{document}
\newtheorem{fig}[figure]{Figure}


\begin{titlepage}
\renewcommand{\baselinestretch}{1}
\renewcommand{\thefootnote}{\alph{footnote}}

\thispagestyle{empty}

\vspace*{-1.3cm} {\bf \hfill LMU--04/97 } 
                
\vspace*{-0.3cm} {\bf \hfill TUM--HEP--273/97 }  
                 
\vspace*{-0.3cm} {\bf \hfill April 97} \vspace*{1.5cm}
{\Large\bf \begin{center} \title \end{center}} \vspace*{1cm}
{\begin{center}
  {\large\sc  \autha\footnote{
  \makebox[1.cm]{Email:} Blumhofer@Photon.HEP.Physik.Uni-Muenchen.DE}}
  \vspace*{0cm}
  {\it \begin{center}
    Sektion Physik \\ Ludwig--Maximilians--Universit\"at
    M\"unchen \\ Theresienstr.37 \\ D--80333 M\"unchen, Germany
  \end{center} }
  {\large\sc \authb\footnote{
  \makebox[1.cm]{Email:} Johannes.Manus@Physik.TU-Muenchen.DE}}
  \vspace*{0cm}
  {\it \begin{center}
    Institut f\"ur Theoretische Physik \\ Technische Universit\"at M\"unchen
    \\ James--Franck--Stra\ss e \\ D--85748 Garching b. M\"unchen, Germany
  \end{center}}
\end{center}}
\vspace*{0cm}
{\Large \bf \begin{center} Abstract \end{center} }
We present a new method solving Schwinger--Dyson equations in gauge
theories via imaginary part analysis. The self--consistency equation for
the fermion self--energy can be reduced from a numerical unstable
eigenvalue problem to a finite dimensional integral which can be
solved easily. We can therefore derive relations between the
dynamically generated fermion mass, the coupling, the gauge boson mass
and the cutoff. For infinite cutoff the fermion mass is independent of
the gauge boson mass which should be the relevant scale. Thus for a
defined fermion mass it is  necessary to incorporate a running
coupling constant. Within our framework this is possible in a well
defined way. The results are reasonable and in agreement with the
renormalization group in the asymptotic regime. 

\renewcommand{\baselinestretch}{1.2}
\end{titlepage}

\newpage
\renewcommand{\thefootnote}{\arabic{footnote}}
\setcounter{footnote}{0}


\section{Introduction}

Almost every theory describing physics beyond the electro--weak scale
incorporates some kind of dynamical scenario. Although symmetries play
an important role in quantum field theory and are well understood,
most of them are broken and the breaking can only effectively be
described. Either the Higgs is a composite particle with a rich
dynamical background or it is replaced by a dynamical model like
technicolor. Even if the Higgs is a fundamental particle, we need some
supersymmetric extension of the standard model to get rid of the
quadratic divergences. But then the SUSY breaking presumably is caused by
some underlying dynamics. Whenever a symmetry is broken a
dynamical mechanism is proposed. One reason might be the fact that up
to now only a running coupling can mediate a symmetry breaking
scenario with a large hierarchy of scales by blowing up or vanishing
at a certain point. 

In contrast our possibilities to deal with dynamical symmetry breaking
are restricted. Most of the statements stem from simple models (e.g.
Nambu--Jona-Lasinio, Gross--Neveu, etc.\cite{Nambu}) or from
renormalization group arguments \cite{BHL}. However the
predictions of dynamical models, e.g. gauge models of top
condensation \cite{King}, require more effort to find appropriate
approximations and reliable numerical analyses. 

In this paper we concentrate on scenarios where dynamical symmetry
breaking is mediated by strongly interacting massive gauge
bosons. The corresponding Schwinger--Dyson equations\footnote{
For an overview see e.g. \cite{Fomin}.} are hard to
solve within a reasonable approximation. One is confronted with a lot 
of problems. Even in ladder approximation the main parameters like the 
critical coupling and the dynamically generated mass of the fermion can 
only be calculated with large uncertainties. The solutions of the 
integral equation strongly depend on the behavior at large momenta. 
The eigenvalue equation thus involves numerical instabilities especially for 
infinite cutoff. Furthermore the integral equations are usually solved 
in the euclidean region. Due to the above mentioned problems and the cut 
in the complex plane it is nearly impossible to make statements about 
the behavior in Minkowski space by analytic continuation.   

One of the key features of dynamical symmetry breaking in asymptotically 
free field theories is that the running coupling defines the scale of
chiral symmetry breaking by dimensional transmutation. However a full
Schwinger--Dyson system with a self--consistent equation for the vertex
is rather complicated and needs some simplifications. Even if one uses
the perturbative renormalization group running for the coupling, the
infrared singularity normally requires assumptions about the behavior
of the coupling in the confinement region.

We present here a new method which solves most of the above
problems. The integral equation is directly treated in Minkowski space,
which simplifies the equation, which is then solvable by integration. This
procedure also enables us to introduce a running coupling in an
elegant way, which is well defined and does not require any
assumptions about the behavior in the nonperturbative infrared region.


\section{The Imaginary Part Method}

We consider Schwinger-Dyson equations in the framework of the ladder
approximation. The self--consistency equations for the vertex
function and the vacuum polarization of the gauge boson are
disregarded. The integral equation for the fermion self--energy is
restricted to single gauge boson exchange and the above mentioned
quantities are substituted by their treelevel values. (A running
coupling will be incorporated in chapter \ref{general}). The resulting
integral equation is often called the massive gap equation:

\begin{minipage}{14.9cm}
\begin{center}
\begin{picture}(230,40)(0,0)
\ArrowLine(20,10)(40,10)        \ArrowLine(50,10)(70,10)
\GOval(45,10)(5,8)(0){0.2}
\Text(95,10)[c]{=}
\ArrowLine(120,10)(145,10)      \ArrowLine(145,10)(195,10)
\ArrowLine(195,10)(220,10)      \PhotonArc(170,10)(25,0,180) 3 7
\GOval(170,10)(5,8)(0){0.7}   
\Vertex(145,10) {1.5}              \Vertex(195,10) {1.5}
\end{picture}
\end{center}
\end{minipage}
\parbox{1.5cm}{\beq \eeq}
  \beq
  iS^{-1}(p) - \slash{p} = \frac{1}{i} \int \frac{d^4 k}{(2 \pi)^4}
  (\Gamma^a \gamma^\mu) S(k) (\Gamma^a \gamma^\nu)
  \frac{-i\left( g_{\mu\nu}-\frac{(p-k)_\mu(p-k)_\nu}{(p-k)^2}\right)}
  {(p-k)^2-M^2} \; .
  \eeq
Here the $\Gamma^a$ includes the gauge part of the coupling. We use
the Landau gauge because the renormalization constant $A(p^2)$ of the
full fermion propagator 
  \beq
  iS^{-1}(p) = A(p^2)\slash{p} - \Sigma(p^2)
  \eeq
is equal to one in that case. With this we get
  \beq
  \Sigma(p^2) = i 3\Gamma^a \Gamma^a \int \frac{d^4 k}{(2 \pi)^4}
  \frac{\Sigma(k^2)}{k^2-[\Sigma(k^2)]^2} \frac{1}{(p-k)^2-M^2} \; .
  \eeq
The physical solution of this integral equation has to vanish for
momenta $p^2\!\!\rightarrow\!\!\infty$, otherwise the integral is
divergent. Therefore the self--energy function cannot be analytical in
the complete complex plane unless it is trivial. This is also reasonable 
since for  momenta 
$p\!>\!\!M\!\!+\!m$ the self--energy function $\Sigma(p^2)$ should have an 
imaginary part because the fermion is then able to emit a real gauge boson. 
We will see this property later in our solutions. For simplicity we use the
denominator approximation 
  \beq
  \Sigma(p^2) = -i C \int\frac{d^4k}{\pi^2}
  \frac{\Sigma(k^2)}{k^2-m^2} \frac{1}{(p-k)^2-M^2}
  \;\;\;\;\mbox{with}\;\;\;\; C=-\frac{3\Gamma^a \Gamma^a}{(4\pi)^2}
  \label{sigmaeq}
  \eeq
where the $\Sigma(p^2)$--function in the denominator is replaced by a
fixed mass $m$. As long as this fermion mass is below the boson mass
$M$ the $\Sigma(p^2)$ in the denominator does not induce any dynamics
and can therefore be replaced\footnote{For $M=0$ a rich dynamical
structure coming from the denominator can appear \cite{Hutter}.}.

Up to that point we have used the conventional procedure. This equation is
usually in the euclidean region where
$\Sigma(p^2)$ is strictly real. In Minkowski space one expects a
coupled system of real and imaginary part, which seems hard to 
solve. But we will see that this is not the case. We can establish a
decoupled integral equation only for the imaginary part, first shown
in \cite{AB}. The real part is then calculable by a dispersion
relation.

First we express $\Sigma(p^2)$ through the dispersion relation
  \beq
  \Sigma(k^2)=-\frac{1}{\pi}\int\limits^\infty_{-\infty}
  \frac{I(\l^2)}{k^2-\l^2+i\varepsilon} \;d \l^2
  \label{dispersion}
  \eeq
where $I(p^2)=\I[\Sigma(p^2)]$ and introduce Feynman parameters.
Angular integration results then in the one-dimensional integral
equation
  \bea
  I(p^2) = C \int\limits_{-\infty}^{\infty} d\l^2
  \, K_I(p^2,\l^2,M^2,m^2)I(\l^2)
  \label{IKI}
  \eea
with the integral kernel
  \bea
  K_I(p^2,\l^2,M^2,m^2) & = &
  \frac{- 2 }{m^2-\l^2} \left[ \theta(p-M-m) \sqrt{\left(
  \frac{p^2-M^2+m^2}{2p^2} \right)^2 - \left( \frac{m}{p} \right)^2 }
  \right. \nonumber \\[0.3cm]
  &   & \left. - \theta(p-M-\l) \sqrt{\left( \frac{p^2-M^2+\l^2}{2p^2}
  \right)^2 - \left( \frac{\l}{p} \right)^2 } \, \right]
  \eea
where $p$ and $\l$ denote the positive square roots of $p^2$ and
$\l^2$.

It can now easily be seen that the imaginary part $I(p^2)$ starts at
\*{M\!\!+\!m}: Let $I(p^2)$ vanish for \*{p^2\!\!<\!\mu^2}. The first part 
of the kernel contributes for \*{p\!\ge\!M\!\!+\!m} and the second part 
for \*{p\!\ge \!\!M\!\!+\!\mu}. Consider \*{\mu\!<\!\!M\!\!+\!m}, then 
only the second part of the kernel contributes for \*{p\!<\!M\!\!+\!m}. 
Hence \*{\mu\!=\!M\!\!+\!\mu} must be true, which is however impossible for 
\*{M\!\!>\!0}. Thus $I(p^2)$ vanishes below \*{M\!\!+\!m}. As mentioned 
above this is exactly the threshold where a real boson can be emitted. 
One should also introduce a cutoff $\Lambda$ to study the dependence on 
the scale of new physics.

Using the well known $\lambda$-function 
  \beq
  \lambda(x,y,z)=\sqrt{x^2+y^2+z^2-2xy-2xz-2yz} 
  \eeq 
and imposing \*{p\!\ge\!M\!\!+\!m} we get
  \beq
  I(p^2)=C\left[\int\limits^{\Lambda^2}_{(M+m)^2}\frac{I(\l^2)}{\l^2-m^2}
  \frac{\lambda(p^2,M^2,m^2)}{p^2}d\l^2-
  \int\limits^{(p-M)^2}_{(M+m)^2}\frac{I(\l^2)}{\l^2-m^2}
  \frac{\lambda(p^2,M^2,\l^2)}{p^2}d\l^2\right] \; .
  \eeq
The first integral is known from the dispersion relation:
  \beq
  \int\limits^{\Lambda^2}_{(M+m)^2}\frac{I(\l^2)}{\l^2-m^2}d\l^2 = m\pi \; . 
  \label{norm}
  \eeq
The integral equation therefore has the form:
  \beq
  I(p^2)=C\left[m\pi\frac{\lambda(p^2,M^2,m^2)}{p^2}-
  \int\limits^{(p-M)^2}_{(M+m)^2}\frac{I(\l^2)}{\l^2-m^2}
  \frac{\lambda(p^2,M^2,\l^2)}{p^2}d\l^2\right] \; .
  \label{I}
  \eeq
Surprisingly this equation is only formally an integral equation.
For \*{M\!\!+\!m\!\!\le\! p\!\!<\!2M\!\!+\!m} the second term in eq.(\ref{I}) 
vanishes and $I(p^2)$ is given analytically by the first term. 
For \*{2M\!\!+\!m\!\!\le \!p\!\!<\!3M\!\!+\!m} $I(p^2)$
is additionally a simple integral of $I(p^2)$ over the already known
region and so on. Thus $I(p^2)$ can be expressed by a finite
dimensional integral and a finite series in $C$. 

This is an important progress in solving Schwinger--Dyson equations
since up to now the integral equations were solved as a rather
unstable eigenvalue problem. The dynamical mass function decreases
rather slowly so that an iteration procedure converges badly. On the
other hand a determination of the eigenvalues of a huge matrix is very
extensive and sensitive to the stepsize.

Nevertheless a determination of $C$ for a fixed cutoff $\Lambda$ is
complicated. One better determines $\Lambda$ as a function of $C$, i.e
one determines $I(p^2)$ up to that scale $\Lambda$ where the relation 
(\ref{norm}) is fulfilled. Further integration to the next points
where $I(p^2)$ fits this normalization requirement yields a whole series 
of possible cutoffs $\Lambda_2,\Lambda_3,\ldots$. This is equivalent to 
different eigenvalues $C$ for a fixed cutoff.


\section{Numerical Solutions}

A numerical analysis of eq.(\ref{I}) is shown in fig.\ref{reim1}
and fig.\ref{reim2}. 
\begin{figure}[ht]
\hspace*{0.25cm}
\bild{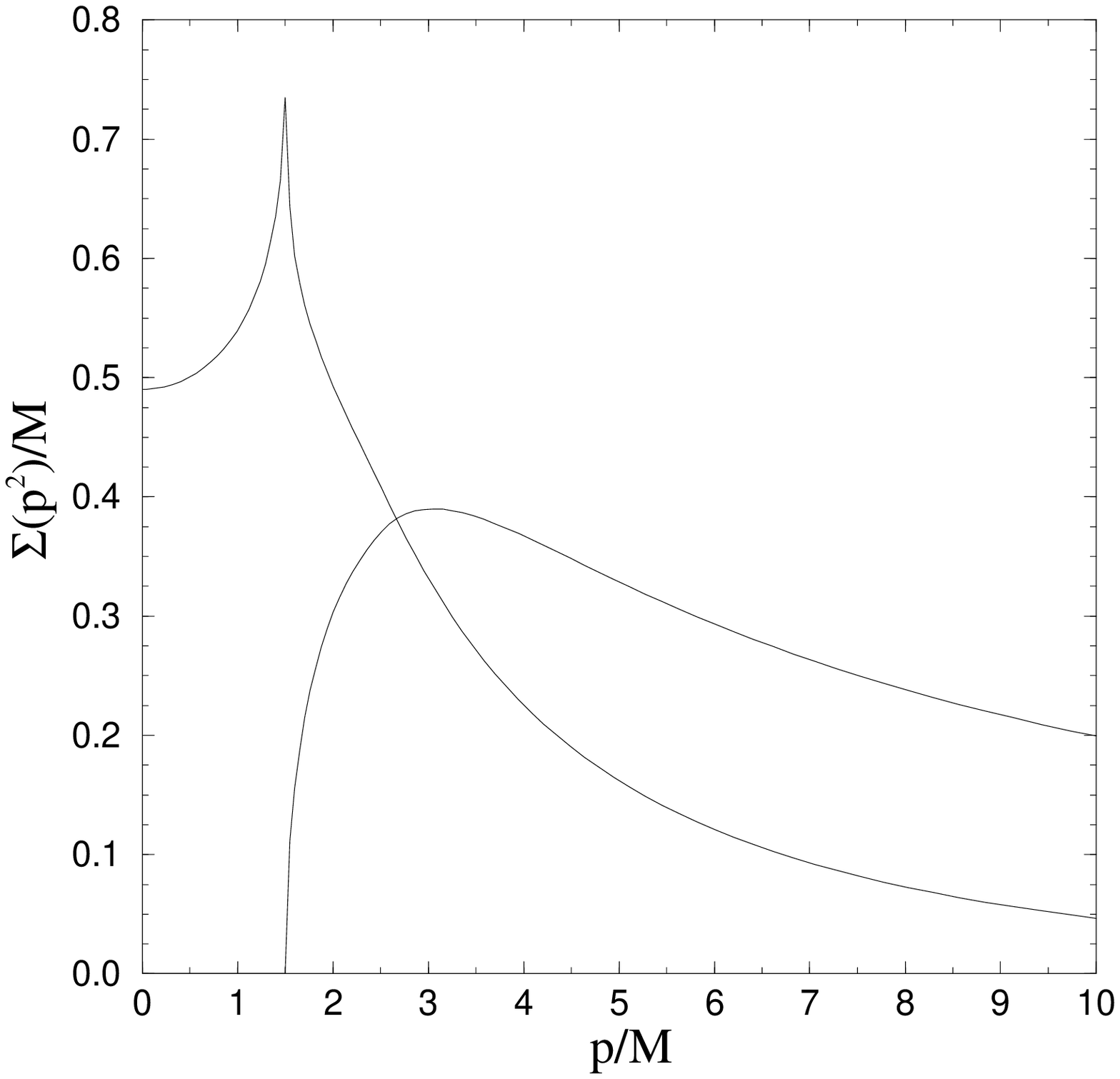}
     {\Text(50,150)[c]{\scriptsize $\R$}
      \Text(110,100)[c]{\scriptsize $\I$} }
     {Real and imaginary part for $m=0.5M$ and $C=0.3$.}
\hspace*{0.4cm}
\bild{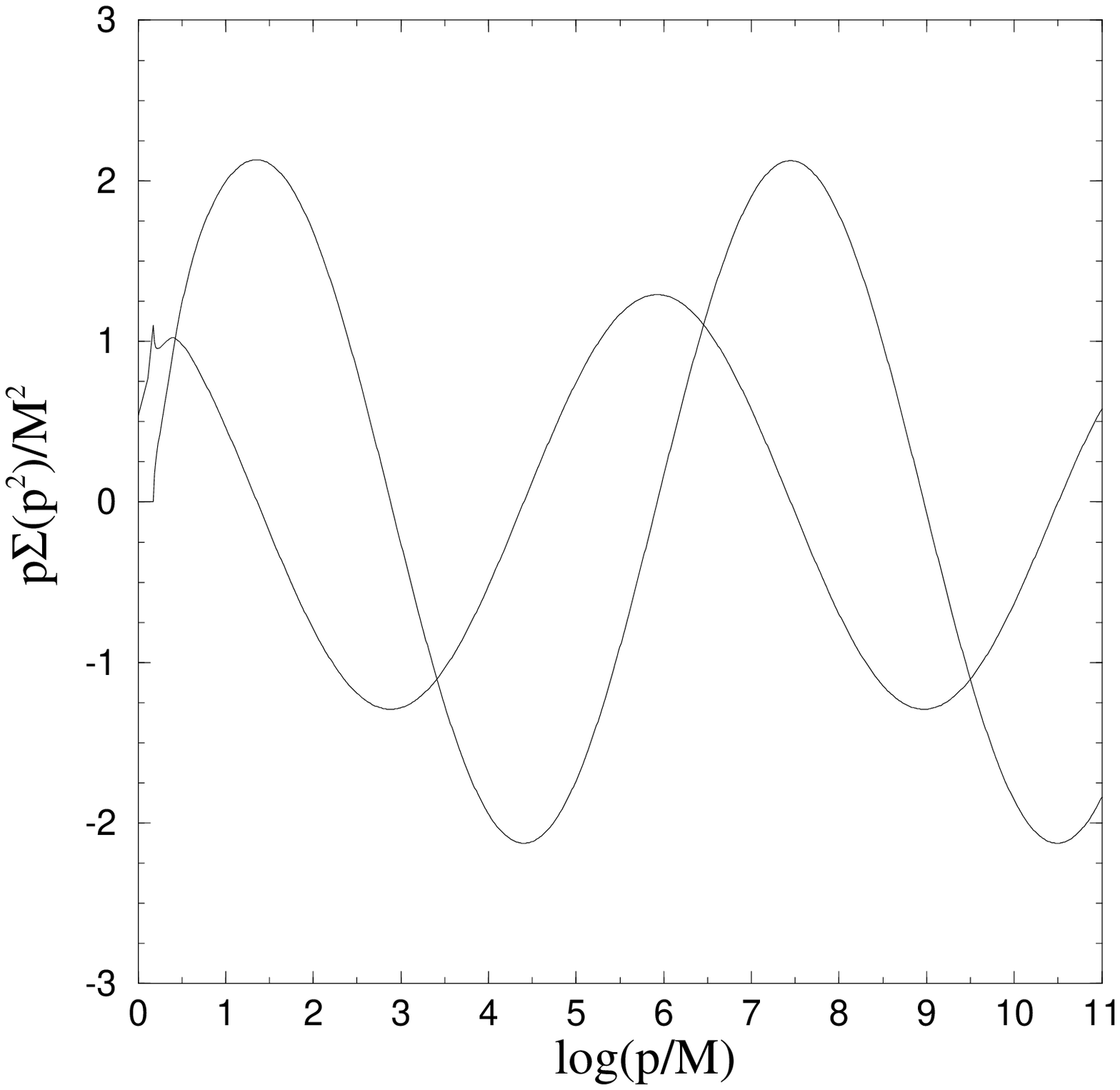}
     {\Text(60,150)[c]{\scriptsize $\I$}
      \Text(30,75)[c]{\scriptsize $\R$} }
     {Asymptotic behavior of the real and imaginary part for $m=0.5M$,
      $C=0.3$ and $\Lambda_4=4 \cdot 10^{11}$.} 
\hspace*{0.25cm}
\end{figure}
We can see that the imaginary part of the self--energy function starts
at the threshold \*{m\!+\!M} as we mentioned before. The real part has 
therefore a ``resonance like'' peak at this point. In fig.\ref{reim2} we
multiplied the self--energy function with the momentum $p$ to compensate
the $1/p$ drop of the solutions and to see the oscillating behavior
in the asymptotic region. We took the fourth cutoff, which we
denoted by $\Lambda_4$ to determine the real part of $\Sigma$. The obtained 
asymptotic behavior is exactly what we
expect for large momenta as it is shown in appendix \ref{app_a}: 
  \beq
  \R\left[\Sigma(p^2)\right],\I\left[\Sigma(p^2)\right]\propto \frac{1}{p}
  \sin\left(\sqrt{C-\frac{1}{4}}\ln\frac{p^2}{\mu^2}+\varphi\right) \; .
  \eeq
This demonstrates the numerical stability of the method over a wide
range of momenta. Additionally we can calculate the cutoff for
different couplings $C$ and masses $m$ and $M$. For each value of $C$
one gets a series of cutoffs $\Lambda_i$, for a fixed cutoff one gets a 
series of eigenvalues $C_i$  respectively, which is shown in fig.\ref{c}. 
Below the critical coupling $C_{krit}=0.25$ dynamical
symmetry breaking does not occur, which was analytically shown by 
Maskawa and Nakajima \cite{Maskawa}. This can now numerically be
confirmed by a regression to the point
\*{\Lambda\!\!\rightarrow\!\!\infty}. This is true for 
all eigensolutions and astonishingly also for different fermion masses
$m$. But this suggests that it is impossible to give a defined fermion
mass depending on the coupling $C$ for infinite cutoff $\Lambda$, which
can be seen in fig.\ref{cm}. The slope of the function $m(C)$ gets
bigger for rising cutoff. For infinite cutoff it is a vertical line,
so the fermion mass is either zero for couplings below the critical
value or infinite for couplings above. Normally one would expect the
fermion mass to be dominated by the mass scale of the gauge
boson. This is however not the case, the fermion mass is shifted up to
the cutoff scale, which corresponds to a hierarchy
problem.\footnote{Thus a serious problem shows up in models where the
dynamical symmetry breakdown is caused by a broken abelian gauge
symmetry. There the coupling is not only constant but rather rises in the
UV region. So the model has to be embedded in some other theory to
remove the Landau pole. The fermion mass is then not only pulled up to
the $U(1)$ gauge boson mass but to the embedding scale.} The coupling
$C$ is large everywhere and a boson mass is not necessary for chiral
symmetry breaking. So why should the fermion mass be small? The cutoff
$\Lambda$ is not at all removed from the problem although the
dynamical fermion mass is finite. We need a running coupling in
addition to get rid of this problem which is the subject of the next
section.
\begin{figure}[t]
\hspace*{0.25cm}
\bild{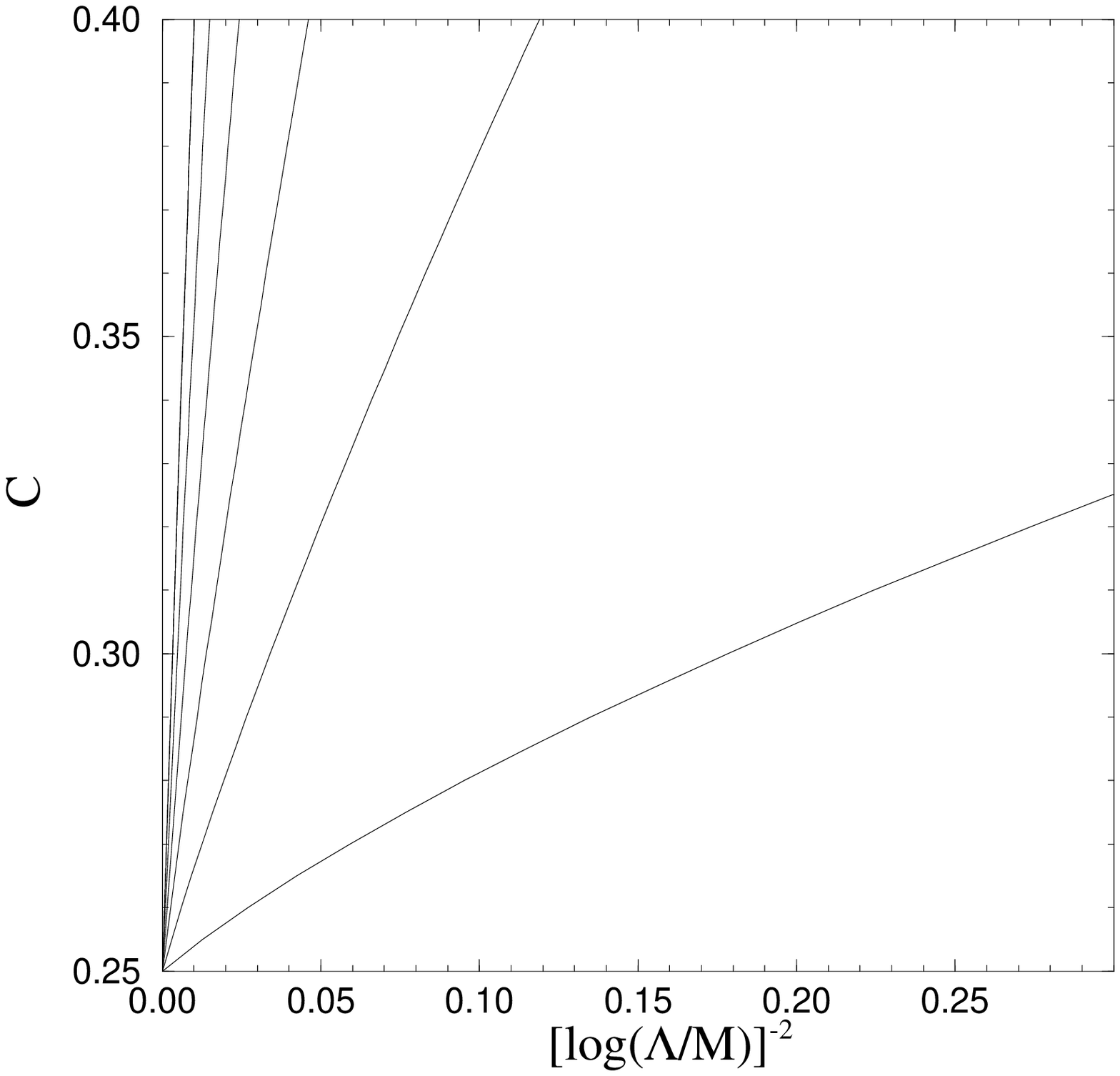}
     { }
     {$C$ for different cutoffs $\Lambda$ with $m=0.1M$.}
\hspace*{0.4cm}
\bild{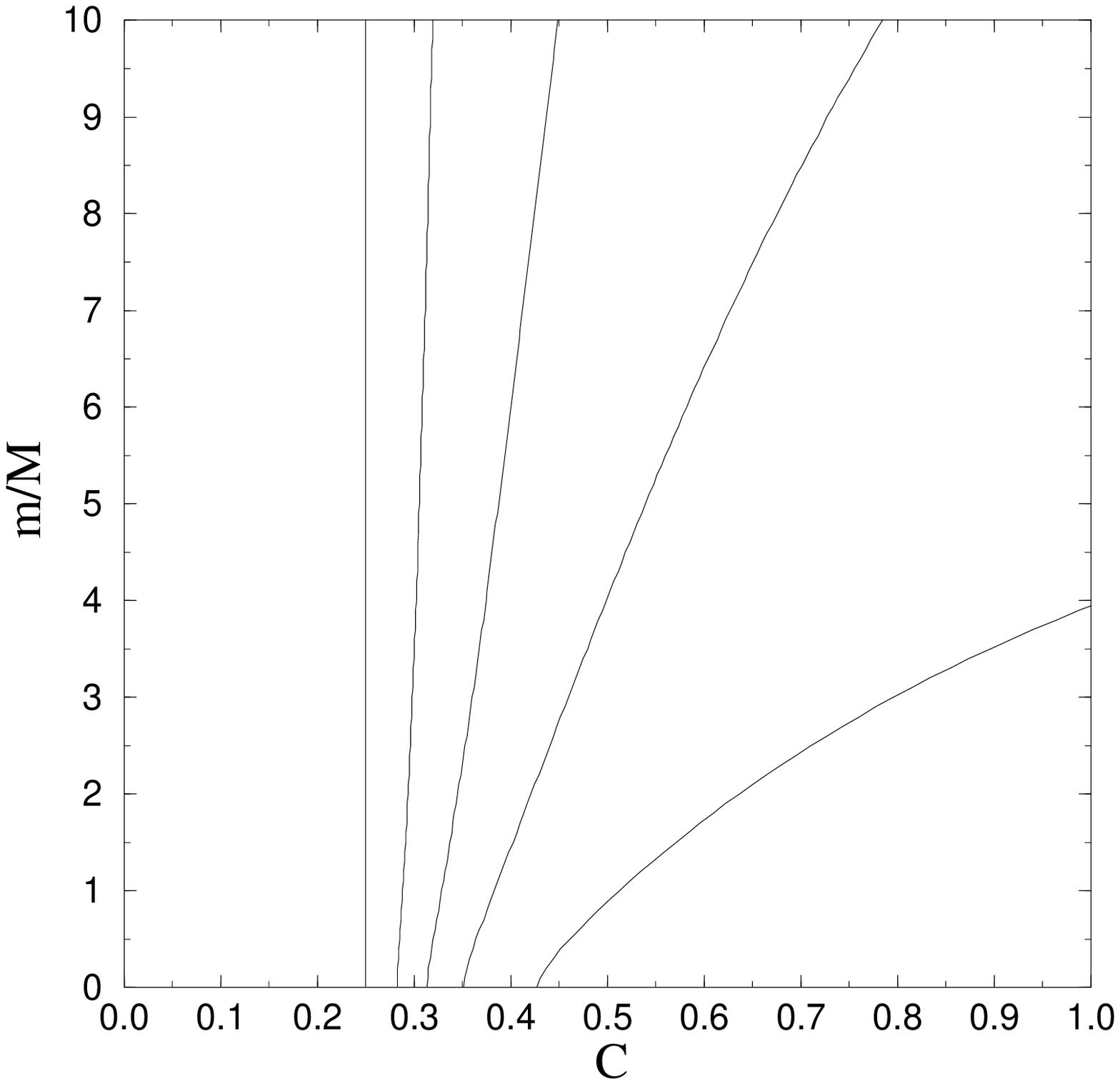}
     {\Text(170,63)[c]{\scriptsize $10$}
      \Text(130,130)[c]{\scriptsize $30$}
      \Text(98,160)[c]{\scriptsize $100$}
      \Text(78,174)[c]{\scriptsize $1000$}
      \Text(35,180)[c]{\scriptsize $\Lambda/M=\infty$}}
     {The C-dependence of $m$ for different cutoffs $\Lambda$.}
\hspace*{0.25cm}
\end{figure}


\section{Generalizations -- The Running Coupling} \label{general}

Let us go back one step and have a look at equation (\ref{IKI})
again. This integral equation has a much deeper meaning.
It can be written in the following form
  \beq
  I(p^2)=\frac{1}{\pi}\int\limits^\infty_0 d\l^2\frac{I(\l^2)}{\l^2-m^2}
  \left( G(p^2,M^2,m^2)-G(p^2,M^2,\l^2) \right)
  \eeq
where
  \beq
  mG(p^2,M^2,m^2) = \I\left( -iC\int\frac{d^4k}{\pi^2}\frac{m}{k^2-m^2}
  \frac{1}{(p-k)^2-M^2} \right)
  \eeq
is the imaginary part of the normal first order perturbation theory.
$G(p^2,M^2,m^2)$ can also be determined via the optical theorem:
  \beq
  G(p^2,M^2,m^2)=\pi C \, \frac{\lambda(p^2,M^2,m^2)}{p^2} \; .
  \eeq
The inner momenta are therefore strictly on shell. This enables us
to introduce a running coupling in an elegant way. The full vertex
depends on the squares of the momenta, i.e. $p^2$, $M^2$ and
$m^2$. Since $p^2\!>\!M^2,m^2$ we can use the mass independent running
coupling as an ansatz for the vertex. In this case $p^2$ is the only
relevant scale which dominates the running of the vertex. The same
arguments can be applied to $G(p^2,M^2,\l^2)$. A running coupling can
therefore be introduced in a very elegant and well controlled
approximation by simply replacing $C$ by $C(p^2)$.\footnote{In the
full system of integral equations only one vertex is substituted by
the full vertex to avoid double counting of diagrams. 
However for the considered one loop diagram one has to substitute both 
vertices since the running coupling does not incorporate the dynamical mass
function.} This appears to be a very trivial transformation,
but the results are not as trivial. A very nice feature of our calculation is
that we do not have to make any assumptions about the coupling below the scale 
\*{M\!+\!m}. Although for an asymptotically free theory
the perturbative coupling develops a pole $\Lambda_{\rm IR}$ in the
infrared, as long as \*{M\!+\!m\!>\!\Lambda_{\rm IR}} we need not care about
the pole in the coupling and the change of the running below the
breaking scale and thus avoid nonperturbative questions in many cases.
We have therefore a kinematical cut shown as dashed line in fig.\ref{rual} and
fig.\ref{excit}.

To determine the running coupling $C(p^2)$ we define the
group theory factors $c_1$, $c_2$ and $c_3$ of a vector gauge theory (in
brackets the values for the fundamental representation of SU$(N)$): 
  \bea
  \begin{array}{ll}
  f_{acd}f_{bcd}=c_1\delta_{ab} \hspace{0.5cm} & (c_1=N) \\[0.2cm]
  Tr(T_aT_b)=c_2\delta_{ab} & \left(c_2=\frac{1}{2}\right) \\[0.2cm] 
  \sum\limits_a T_aT_a=c_3 \mbox{1\hspace{-1.2mm}l} & 
  \left(c_3=\frac{N^2-1}{2N}\right) \mbox{.} 
  \end{array}
  \eea
With this definitions we get
  \beq
  C(p^2)=\frac{3c_3}{(4\pi)^2}g^2(p^2) \; .
  \eeq
The one--loop running coupling can be written in the form
  \beq
  g^2(p^2)=\frac{(4\pi)^2}{\beta_0 \ln\frac{p^2}{\Lambda^2_{\rm IR}}}
  \hspace{1cm} \mbox{with} \hspace{1cm}
  \beta_0 = \frac{11}{3}c_1-\frac{4}{3}\sum\limits_R c_2^R \; .
  \eeq
Thus we obtain
  \beq
  C(p^2)=\frac{\kappa}{\ln\frac{p^2}{\Lambda^2_{\rm IR}}} \hspace{1cm}
  \mbox{with} \hspace{1cm} \kappa=\frac{3c_3}{\beta_0} \; .
  \label{cvpq}
  \eeq

\begin{figure}[t]
\hspace*{0.25cm}
\bild{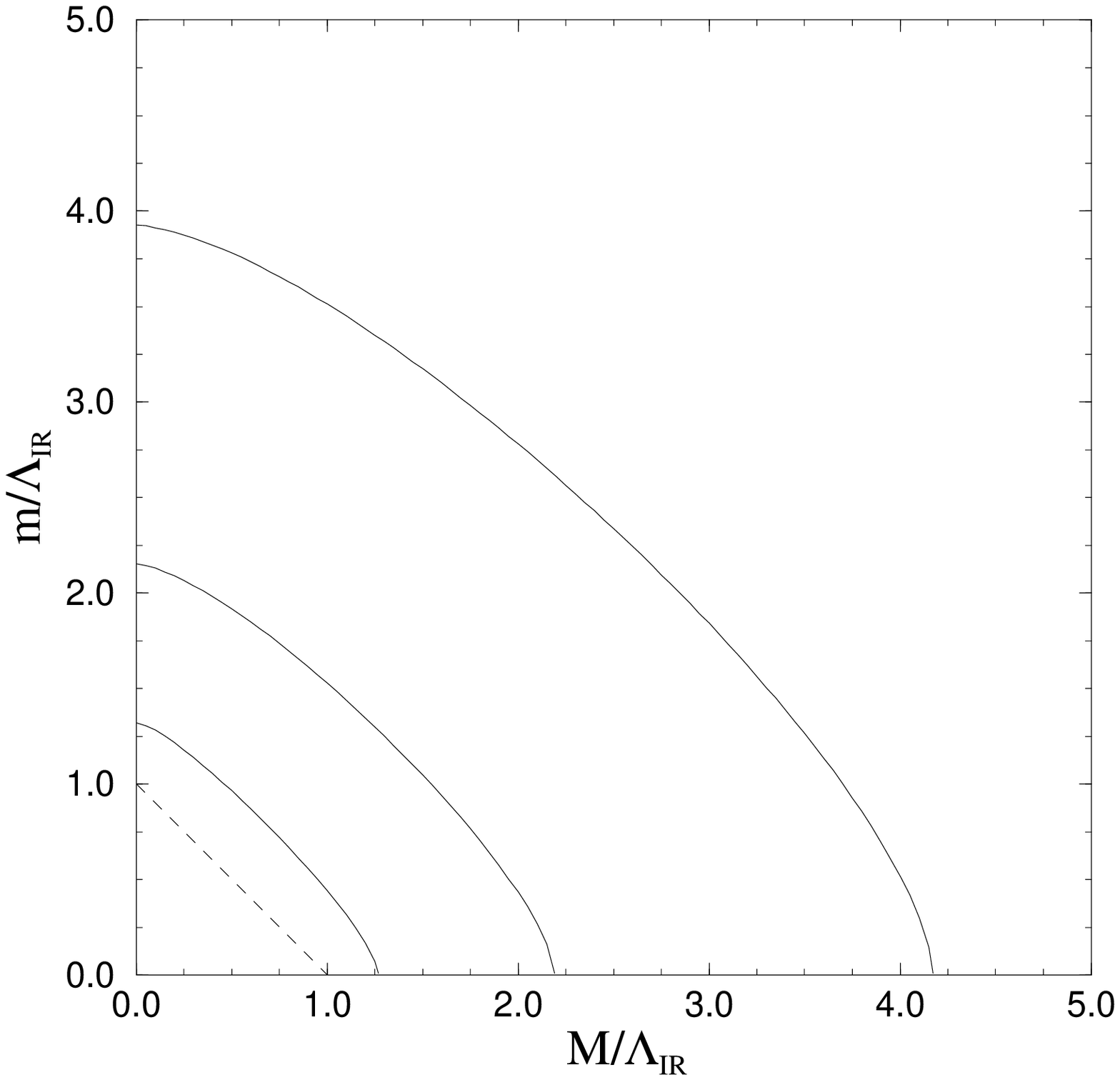}
     {\Text(110,110)[c]{\scriptsize $\kappa=2$}
      \Text(70,65)[c]{\scriptsize $\kappa=1.5$}
      \Text(45,42)[c]{\scriptsize $\kappa=1$}}
     {$m$ for different boson masses with $C(p^2)=
     \kappa\left( \ln\frac{p^2}{\Lambda_{\rm IR}^2}\right)^{-1}$, 
     $\kappa=1,1.5,2$.}
\hspace*{0.4cm}
\bild{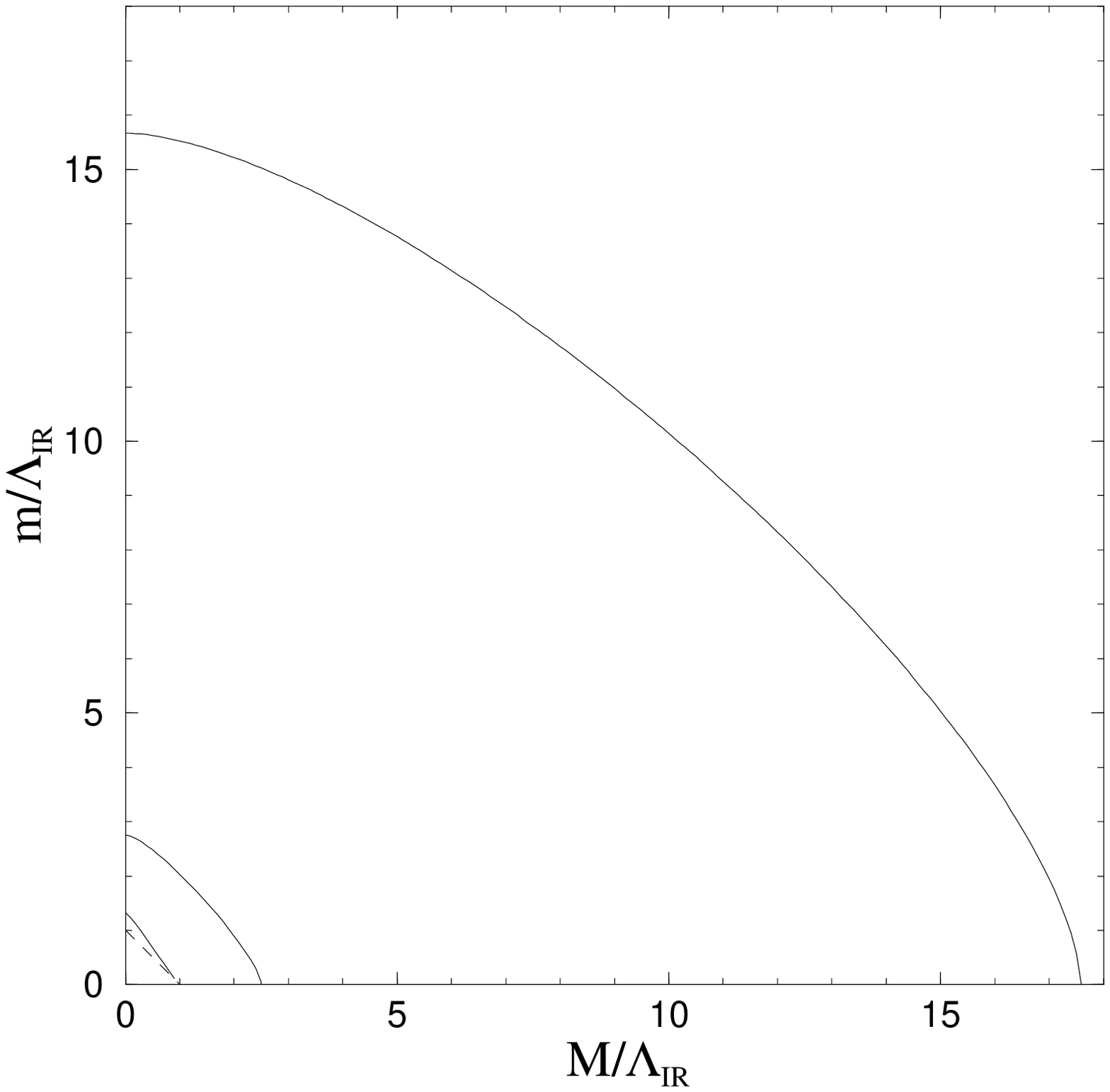}
     { }
     {$m$ for different boson masses and $\kappa=3$ with excited 
     states.}
\hspace*{0.25cm}
\end{figure}

In fig.\ref{rual} the numerical results for different $\kappa$--values
are shown in the limit $\Lambda\!\!\to\!\!\infty$. The dynamically
generated fermion mass depends on the ratio $M/\Lambda_{\rm IR}$.
If the boson mass is bigger than a certain scale, the fermion remains
massless. This corresponds to a critical value of the coupling at the
breaking scale. If $\kappa$ is smaller than $\approx\!1$ and
$M\!\!<\!\Lambda_{\rm IR}$ the imaginary part gets large contributions from
the coupling around the pole at $\Lambda_{\rm IR}$. Even if $\kappa$
goes to zero a dynamical mass $m$ should unphysically be generated
with $m\!>\!\!M\!\!-\!\Lambda_{\rm IR}$. Because of the logarithmic pole
structure of the coupling the difference $m\!-\!\Lambda_{\rm IR}$ goes very
fast to zero so that we cannot calculate $m$ for very small values of
$\kappa$. There one should thus replace the pole of the coupling by a more
smoothly running coupling which tends to a finite value at zero. 
The \*{m-\kappa} dependency is shown in fig.\ref{kam} and \ref{kam2}. For big
values of $\kappa$ the fermion mass $m$ depends logarithmically on $\kappa$. 

SU$(N)$ gauge theories give $\kappa \gapprox \frac{9}{22} \approx 0.41$ for 
large $N$ and are therefore just in the valid region. Theories with large 
fermion representations like QCD with colorsextet quarks or
technicolor can have even larger fermion masses than $\Lambda_{\rm
IR}$ because of a larger $\kappa$--value. For QCD including only the
three lightest quarks u,d,s, which have current masses smaller than
the infrared pole, we get $\kappa=\frac{4}{9} \approx 0.44$. Since $M\!=0$ we
obtain a ``constituent mass'' around $\Lambda_{\rm QCD}$\footnote{For
the average values of the current quark masses and $\alpha_s=0.118$ one gets
$\Lambda^{(3)}_{\rm QCD}=327$ MeV. (For the used three-loop beta-function see
\cite{PDG})}. 

Note that we do not need any further input parameters to determine
this relation between $\Lambda_{\rm IR}$ and the constituent quark
mass provided that $\kappa$ is large enough. It is calculated from
first principles and only depends on the approximation.

As in the case without running coupling, we can get more than one
solution, but now a finite number of solutions (See fig.\ref{excit}).
These ``excited solutions'' are particle excitations related to different
vacuum states with smaller masses and can therefore only appear in a
bubble of the corresponding vacuum. It is not clear, whether there is any
physical relevance of these additional solutions. However if one plays
with this consideration, as it is done in \cite{Triantaphyllou}, the
running coupling provides the possibility to have a finite
spectrum, which is controlled by $M$ and $\Lambda_{\rm IR}$.

\begin{figure}[t]
\hspace*{0.25cm}
\bild{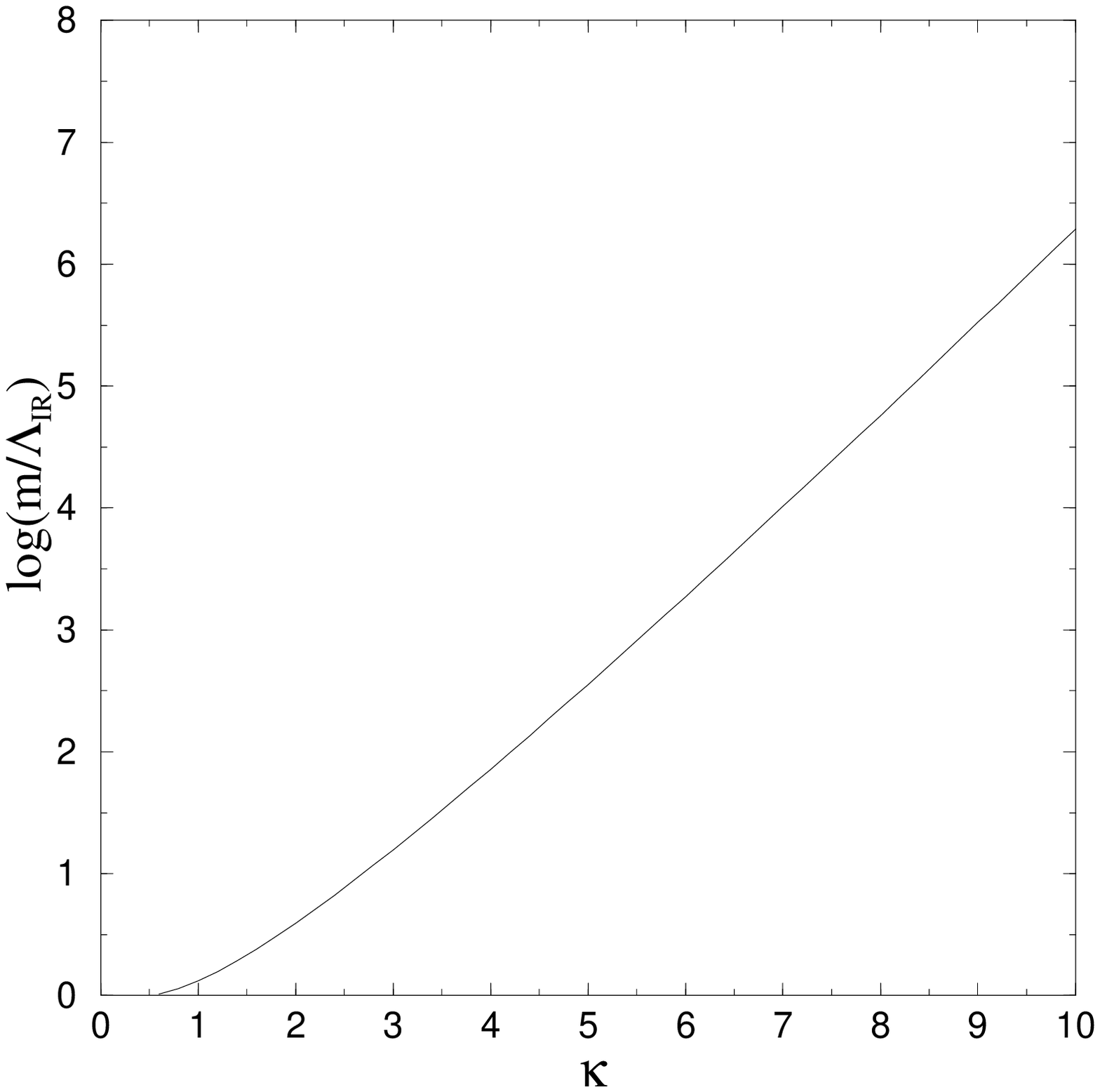}
     { }
     {$\log\left(\frac{m}{\Lambda_{\rm IR}}\right)$ for different
     $\kappa$-values and $M=0$.}
\hspace*{0.4cm}
\bild{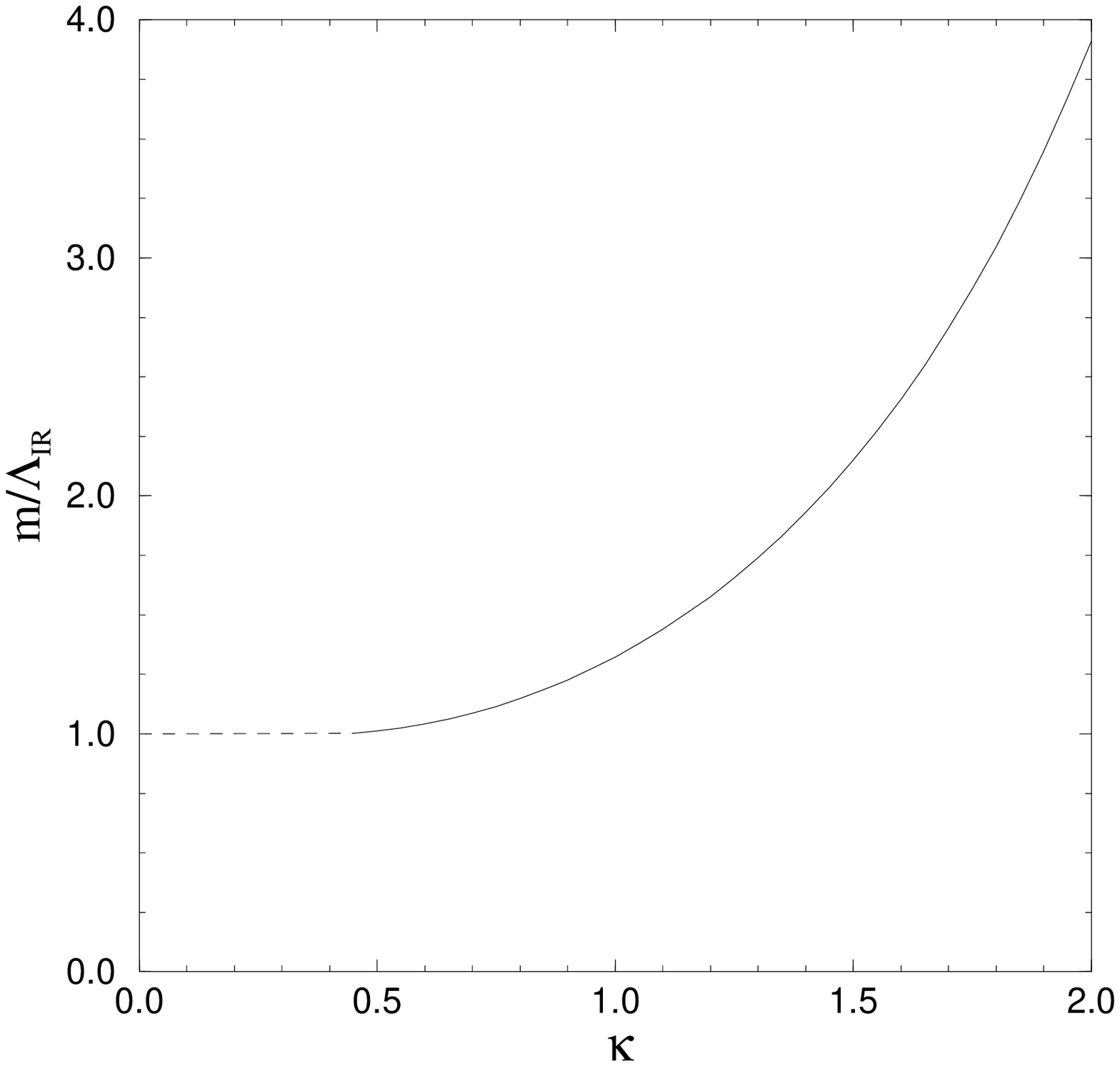}
     { }
     {$\frac{m}{\Lambda_{\rm IR}}$ for small values of $\kappa$ and
      $M=0$. For the dashed line see text.}
\hspace*{0.25cm}
\end{figure}

As in the non running case we can compare the asymptotic of our
solution, now with the asymptotic which is given by the
renormalization group. One expects for the irregular and regular part of
the solution (\cite{Pagels} and appendix \ref{app_b})
  \beq
  \Sigma_{\rm irreg}\propto\frac{1}{\left(\ln \frac{p}{\Lambda_{\rm IR}}
  \right)^\kappa} \;\;\;\; \mbox{and} \;\;\;\;
  \Sigma_{\rm reg}\propto\frac{\left(\ln\frac{p}{\Lambda_{\rm IR}}\right)
  ^\kappa}{p^2} \; .
  \label{reas}
  \eeq
A fit to the asymptotic part of our solutions reproduces the
anomalous dimension within a few percent. We only see the regular
solution up to scales of the cutoff. This is clear, because we did not
introduce a current mass up to now. This follows in the next chapter.


\section{Schwinger--Dyson Equation with a Current Mass}

So far we have assumed that the integral
  \beq
  \int\limits^{\Lambda^2}_{(M+m)^2}\frac{I(\l^2)}{k^2-\l^2+i\varepsilon} 
  \; d\l^2
  \label{integral}
  \eeq
is convergent. In that case we have automatically obtained only
regular solutions 
  \beq
  \Sigma_{\rm reg}\propto\frac{(\ln p)^\kappa}{p^2} \; .
  \eeq
This is consistent with the assumption since the integral
(\ref{integral}) is convergent. If it diverges the
dispersion relation (\ref{dispersion}) is not correct. We need then in
addition the integral over a half circle in the complex plane which
compensates the divergence. This infinite constant can be replaced by
an integral along the real axis as it is done for subtracted
dispersion relations
  \beq
  \Sigma(k^2)=-\frac{1}{\pi}\int\limits^\infty_{-\infty}
  \frac{I(\l^2)}{k^2-\l^2+i\varepsilon} \; d\l^2
  +\frac{1}{\pi}\int\limits^\infty_{-\infty}
  \frac{I(\l^2)}{-\mu^2-\l^2+i\varepsilon} \; d\l^2+m_\mu \; .
  \label{dispersion2}
  \eeq
The last two terms are arranged in such a way that
$\Sigma(-\mu^2)\!=\!m_\mu$. Thus we have the freedom to fix the mass function
$\Sigma$ at a certain euclidean\footnote{Since the fermion mass is getting 
complex beyond the threshold it is useful to define it at an euclidean
scale.} scale $\mu$ and therefore to choose a relatively
large fermion mass at scales far beyond the dynamical symmetry breaking
scale. This corresponds to having a bare mass, which is the sum of the
last two terms in eq.(\ref{dispersion2}) and is, of course, infinite.
Putting eq.(\ref{dispersion2}) into eq.(\ref{sigmaeq}) leads to
the same equation (\ref{I}). However instead of eq.(\ref{norm}) we
have another cutoff--mass relation
  \beq
  m\pi=\int\limits_{(M+m)^2}^{\Lambda^2}
  \left(\frac{1}{l^2-m^2}-\frac{1}{l^2+\mu^2}\right)I(\l^2) 
  \; d\l^2+m_\mu\pi \; .
  \label{norm2}
  \eeq
Using the solution $I(p^2)$ for given masses one can calculate the
cutoff $\Lambda_0$ from eq.(\ref{norm}), which is smaller than the
cutoff $\Lambda_{m_\mu}$ from eq.(\ref{norm2}). Between both cutoffs
$I(p^2)$ has an irregular running which corresponds to the renormalization
group running. In the case of a bare mass, the cutoff $\Lambda_0$
corresponds therefore to the scale where chiral symmetry breaking occurs.


\section{Conclusion}

In this work we have presented a new method solving Schwinger--Dyson 
equations in ladder approximation. We have established the corresponding 
integral equation for the imaginary part of the dynamical mass function, 
which is no more an eigenvalue problem but can be solved by integration. 
The procedure is therefore numerically stable. The perfect agreement of
the shape to the analytic solution in the asymptotic regime is an 
impressive verification for this.
We have calculated mass - coupling relations for 
the non-running coupling case, which states whether there is a finetuning or
not. If the cutoff goes to infinity the fermion mass also diverges in
the broken phase and thus 
no defined value is predicted. This is even worse than the usual finetuning 
problem in Nambu--Jona-Lasinio models between the fermion mass and the Fermi 
scale. If the contact interaction represents the exchange of a massive gauge 
boson the Fermi scale is no longer the cutoff of the gauge theory but is 
connected to the gauge boson mass. In the full theory the fermion mass is not
shifted to the Fermi scale but to the cutoff. The only way out of this
dilemma seems to be the introduction of a running coupling within an
asymptotically free theory. We have shown that in our framework this is
possible in a well defined way. Now the fermion mass is calculable in
dependency of $M$, $\Lambda_{\rm IR}$ and $\kappa$ and for $\Lambda
\!\! \rightarrow \!\! \infty$. These are only very
fundamental constants of a gauge theory and in principle no further
assumptions are necessary. Our results are in good agreement with the
common knowledge about dynamical symmetry breaking, only for very
small values of $\kappa$ more assumptions seem to be necessary. This
should be a point of further investigations. The application to many
questions in models of electro--weak symmetry breaking like e.g.
technicolor, topcolor or combined models seems now to be possible with
the here developed tools.

\vspace{.5cm}
{\bf Acknowledgments:} 
We would like to thank Richard Dawid, Dimitris Kominis 
and Manfred Lindner for inspiring and clarifying discussions.

This work was supported in part by the DFG under contract Li519/2-2.

\newpage


\vspace{1cm}
\begin{appendix}

{\LARGE \bf Appendix} \vspace{-0.7cm}
\section{Asymptotic Properties} \label{app_a}

The asymptotic properties of the Schwinger--Dyson equation
(\ref{sigmaeq}) are identical to the massless limit $M\!\!=\!0$ since the 
boson mass $M$ has no influence on the general shape of the solution at 
large momenta. In that limit the integral equation can be transformed to
the differential equation
  \beq
  \left(\frac{d}{dp^2}\right)^2\left(p^2\Sigma(p^2)\right)=
  -C\frac{\Sigma(p^2)}{p^2-m^2}
  \label{dgl}
  \eeq 
which has the solution \cite{Miranskii}
  \beq
  \Sigma(p^2)=m\cdot \left._2F_1\right.\left(\frac{1}{2}+\sqrt{\frac{1}{4}-C},
  \frac{1}{2}-\sqrt{\frac{1}{4}-C};2;\frac{p^2}{m^2}\right) \; .
  \eeq
Here $\left._2F_1\right.$ is the hypergeometric function. Its
asymptotic can also be derived directly from the differential equation
(\ref{dgl}) by neglecting $m$ in the denominator and using the ansatz 
$\Sigma\propto(p^2)^{-\delta}$. One finds
  \beq
  \delta^2-\delta+C=0 \;\;\;\; \mbox{or} \;\;\;\; 
  \delta=\frac{1}{2}\pm\sqrt{\frac{1}{4}-C} \; . 
  \eeq
Only if $C$ is bigger than $1/4$ the solution of the differential equation 
also fulfills the integral equation. Otherwise the boundary conditions 
cannot be fulfilled. Hence $\delta$ is complex.
The real and imaginary part of the asymptotic has therefore the form
  \beq
  \R\left[\Sigma(p^2)\right],\I\left[\Sigma(p^2)\right]\propto \frac{1}{p}
  \sin\left(\sqrt{C-\frac{1}{4}}\ln\frac{p^2}{\mu^2}+\varphi\right) \; .
  \eeq

\section{Asymptotic Properties for a Running Coupling} \label{app_b}

In the case $M=0$ our integral equation has the form
  \beq
  I(p^2)=C(p^2)\left(m\pi\frac{p^2-m^2}{p^2}-\int\limits_{m^2}^{p^2}d\l^2
  \frac{I(\l^2)}{\l^2-m^2}\frac{p^2-\l^2}{p^2}\right)  
  \eeq
which can be transformed to the differential equation:
  \beq 
  \left(\frac{d}{dp^2}\right)^2
  \left(I(p^2)\frac{p^2}{C(p^2)}\right)=-\frac{I(p^2)}{p^2-m^2} \; . 
  \eeq  
Using the running coupling (\ref{cvpq}), we make the ansatz
  \beq
  I(p^2)\propto (p^2)^{-\delta}\left(\ln\frac{p^2}{\Lambda_{\rm IR}}\right)^
  {-\gamma}
  \eeq
to describe the asymptotic properties. 
We consider the leading terms in $p^2$ and $\ln\frac{p^2}{\Lambda_{\rm
  IR}}$\
and find the relations $\delta(1-\delta)=0$ and $(1-2\delta)(1-\gamma)=
-\kappa$, which give the two solutions
  \beq
  I(p^2)\propto\frac{1}{\left(\ln\frac{p^2}{\Lambda_{\rm IR}}\right)^{\kappa+1}}
  \;\;\;\; \mbox{and} \;\;\;\; 
  I(p^2)\propto\frac{\left(\ln\frac{p^2}{\Lambda_{\rm IR}}\right)
  ^{\kappa-1}}{p^2}
  \eeq
which correspond to the real counterparts (\ref{reas}).    
\end{appendix}


\parskip=0ex plus 1ex minus 1ex

\end{document}